
\magnification=\magstep1
\hsize 32 pc
\vsize 42 pc
\baselineskip = 24 true pt
\centerline{\bf Macroscopic Equation of Motion
 in Inhomogeneous media:}
\centerline{\bf a Microscopic Treatment}
\vskip 1.0 true cm
\centerline{A.M. Jayannavar and Mangal C. Mahato}
\centerline{Institute of Physics, Sachivalaya Marg,}
\centerline{Bhubaneswar - 751 005, INDIA}
\vskip 1.0 true cm
\noindent{\bf Abstract}
\vskip 1.0 true cm

	The dynamical evolution of a Brownian particle in an
inhomogeneous medium with spatially varying friction and
temperature field is important to understand conceptually. It
requires to address the basic problem of relative stability of
states in nonequilibrium systems which has been a subject of
debate for over several decades. The theoretical treatments
adopted so far are mostly phenomenological in nature. In this
work we give a
microscopic treatment of this problem. We derive the Langevin
equation of motion and the associated Fokker-Planck equation.
The correct reduced description of the Kramers equation in the
overdamped limit (Smoluchowski equation) is obtained. Our
microscopic treatment may be helpful in understanding the
working of thermal ratchets, a problem of much current interest.
\vskip .2 true cm
{\bf Key words:} Brownian particle, diffusion, inhomogeneous
systems, relative stability of states, Fokker-Planck equation.
\vskip .2 true cm
{\bf PACS Numbers:} 05.60.+w, 05.40.+j, 05.90.+m, 02.50.Wp
\vfill
\eject
\vskip .3 true cm
\noindent{\bf 1. Introduction}
\vskip .2 true cm
	Thermodynamic equilibrium states are ideal limiting
cases and are convenient and often theoretically amenable to study with
relative ease. However, one commonly encounters systems that are
away from equilibrium. All nonequilibrium systems relax
naturally toward their respective equilibrium or stationary
states. In nature, evolution is an ongoing and dominant
process. Naturally, the process of relaxation of the
nonequilibrium systems is of great interest in all branches of
natural science, be it physics, chemistry or biology. Moreover,
one comes across nonuniform systems more often than uniform
systems. Uniform systems are characterized by a constant (space
independent) diffusion coefficient throughout the system and
having the same temperature in all parts of the system. There
are well-established theoretical formalisms for uniform systems
to describe their evolution towards equilibrium or steady
states[1].  However, the same is not true for nonuniform systems.
There exist phenomenological descriptions but often without
microscopic foundations. The ad-hoc nature of these desciptions
 have led to some
controversies, too, in the past. For instance, should the
diffusion equation of a Brownian particle in the absence of
external potential have the form[2-8]
$${\partial P\over\partial t} = {\partial^2\over\partial q^2} D(q)P$$
or
$${\partial P\over\partial t} = {\partial\over\partial q} D(q)
{\partial P\over \partial q}  ?$$
However, considerable progress has been made in the last decades
to make the theories of relaxation of nonuniform systems
self-consistent[2,4,8,10].

	In this work, we seek to clarify some of the issues
pertaining to this important case of nonuniform systems in a
systematic manner. We derive, from microscopic theory, the
Kramers equation for the joint probability distribution of position and
velocity of a Brownian particle in an inhomogeneous,
nonisothermal medium. We then proceed to find the correct
Smoluchowski limit to the Kramers equation. This, however, is
not a mathematical problem alone; the underlying conceptual
development is quite appealing and, as mentioned earlier, is
subject to ongoing controversies for over several decades[2-8].

	The evolution of a Brownian particle in condensed media is
the most familiar example of a nonequilibrium process. The process
is accompanied by frictional dissipation but aided by
associated fluctuations. Nonequilibrium behaviour of macroscopic
uniform systems is described well by linear-response theory when the
initial state of the system is close to equilibrium. The
fluctuation-dissipation theorem relating the power absorbed by
the system to the intrinsic fluctuations in the system in
equilibrium has foundations in the linear-response theory.
But, when the system is far from equilibrium the linear response
theory cannot be relied upon. In most of the physical systems,
whether close to or far from equilibrium, the approach to
equilibrium can, however, be liked to one kind or other of a
diffusion process; it may be translational diffusion of
particles, rotational diffusion of macromolecules, spin diffusion
of spin systems, heat or thermal diffusion in solids, energy
diffusion in excitonic motion in semiconductors, and so on. The
diffusion process in inhomogeneous systems, therefore, calls for
added attention.

	In the case of uniform systems the diffusion constant $D$
$= \eta^{-1} k_BT$, where $\eta$ is the friction coefficient and $T$
the temperature. However, for nonuniform systems the space, $q$,
dependence of the diffusion comes separately through $\eta(q)$ and
$T(q)$. The origin of $\eta(q)$ and $T(q)$ and the manner in which
they influence the relaxation of the nonequilibrium system are
entirely different. The variation of $\eta(q)$ (in the absence of
spatial variation of temperature) influences the dynamics of
particle in a potential field and helps the system to approach
towards its equilibrium or steady states. The relative stability of
the competing states is generally governed by the usual
Boltzmann factor in the local neighbourhood of the corresponding
(representative) potential wells. A change in the potential
barrier between two potential well minima changes the relaxation
rate but leaves the relative stability of the two well-states
unchanged. This simple fact, however, may not apply for more
general systems when the temperature is nonuniform
along the potential surface (or spatial coordinate).

	Landauer, in a series of papers[2-5], argues that for systems
with nonuniform temperature the relative stability of two states
will be affected by the detailed kinetics all along the pathways
(on the potential surface) between the two states under comparison. It
is the effect of thermal fluctuations that plays a crucial role
and the resulting effective potential surface may have
completely different nature from that with uniform temperature.
With the help of his "blowtorch" theorem Landauer[3] shows that a
change of temperature away from uniformity even at very unlikely
positions of the system on the potential surface may cause
probability currents to set in moving the system towards a new
steady state situation changing thereby the relative stability of
the otherwise locally stable states. This known important
fact, however, has received much less attention in the
literature than it deserves. This effect can have important
consequences on the particle motion in nonuniform systems, for
instance, the kinetics of growth of crystalline nuclei in the
melt around its critical size. The latent heat generation being,
in this example, responsible for the creation of nonuniform
temperature field across the surface of the nucleus. Nonuniform
temperature field can also be generated by shining light on
semiconductors. One can have nonuniform temperature field also
because of nonuniform distribution of electrons and of phonons
(or of quasiparticles in general)
with different characteristic temperatures
in a solid. It has been suggested that the nonuniform
temperature field can produce current in a closed ring[3,6,9]. There
has been a lot of theoretical work reported in recent times on
thermal ratchets[10]. These works are inspired by the
observed predominantly unidirectional
protein (macromolecule) motion in biological systems even in the
absence of obvious external forces and thermal gradients.
The idea of relative stability of states
in nonuniform temperature systems can help to understand the
working of the thermal ratchets better[11, 12]. These are but few examples
where nonuniform temperature field can have important
bearing on the dynamical evolution. A systematic formalism to deal with such a
situation is, therefore, essential.

	In the following sections we proceed systematically to
set up a formalism from microscopic theory. We derive the
Kramers equation for space dependent friction coefficient and
nonuniform temperature field. We then go over to obtain the
correct Smoluchowski limit of the Fokker-Planck equation. Before
concluding we also give the correct Langevin equation in the
overdamped limit that is approximated properly to order
$\eta(q)^{-1}$.
\vskip .3 true cm

\noindent{\bf 2. Microscopic derivation of Langevin equation in a space
dependent friction field}
\vskip .2 true cm

	To obtain Langevin equation in a space dependent
friction field we
consider the motion of a subsystem (Brownian particle) described
by its cordinate $Q$ and momentum $P$ and subjected to an external
potential field $V(Q)$ of the system. We assume the subsystem to
be in contact with a thermal (phonon) bath. The bath oscillators
are described by coordinates $q_\alpha$ and momentum $p_\alpha$
with characteristic frequencies $\omega_\alpha$. For our
calculation we consider the total Hamiltonian
$${\cal H} = {P^2\over 2M}+ V(Q) +\sum_{\alpha} \bigg [
{p^2_{\alpha}\over 2m_{\alpha}}+{m_{\alpha}\omega^2_{\alpha}\over 2}
(q_{\alpha}-\lambda_{\alpha} {A(Q)\over
m_{\alpha}\omega^2_{\alpha}})^2\bigg ],\eqno{(1)}$$
where $M$ is the Brownian particle mass and $m_\alpha$ are the
masses of the bath oscillators. The interaction of the subsystem
with the thermal bath[13] is through the linear coordinate-coordinate
 coupling term $\lambda_{\alpha} q_{\alpha}A(Q)$. From eq.
(1) one obtains the following equations of motion.
$$ \dot Q = {P\over M},\eqno{(2a)}$$
$$\dot P = - V'(Q)+\sum_{\alpha} \lambda_{\alpha}A'(Q)\bigg [
q_{\alpha}-\lambda_{\alpha}{A(Q)\over m_{\alpha}\omega^2_{\alpha}}
   \bigg ],\eqno{(2b)}$$
$$\dot q_{\alpha} = {p_{\alpha}\over m_{\alpha}},\eqno{(2c)}$$
and
$$\dot p_{\alpha} = -m_{\alpha} \omega^2_{\alpha} q_{\alpha} +
\lambda_{\alpha}A(Q),\eqno{(2d)}$$
where $A'(Q)$ is the derivative of $A(Q)$ with respect to $Q$. After
solving (2c) and (2d) for $q_\alpha$  using the method of
Laplace transform and substituting its value
in (2b), we obtain the Langevin equation of motion for $Q$
and $P$.
$$\dot Q = {P\over M},\eqno{(3a)}$$
$$\dot P = - V'(Q) - \eta [ A'(Q)]^2 {P \over M} + A'(Q) f(t).\eqno{(3b)}$$
Thus, the effect of interaction of the Brownian particle with
the thermal bath is to introduce a friction term and a
fluctuating term $f(t)$ in the equation of its motion. The
fluctuating term is given by
$$f(t) =\sum_{\alpha}\lambda_{\alpha}\bigg [ q_{\alpha} (0) cos
(\omega_{\alpha} t) + {\dot q_{\alpha}(0)\over \omega_{\alpha}} sin
(\omega_{\alpha}
t)\bigg ],\eqno{(4a)}$$
where $q_{\alpha}(0)$ and $\dot q_{\alpha}(0)$ are the initial
positions and velocities of the bath variables. The force $f(t)$ is
fluctuating in character because of the associated uncertainties
in these initial conditions of the bath variables. However, as
the thermal bath is characterized by its temperature $T$, the
equilibrium distribution of bath variables is given by the Boltzmannian form,
so that $f(t)$ follows the following statistics.
$$ < f(t) > = 0\eqno{(4b)}$$
and
$$\eqalign{ <f(t)f(t') > & =\sum_{\alpha} {\lambda^2_{\alpha}k_B T\over
m_{\alpha}\omega^2_{\alpha}} cos (\omega_{\alpha} (t-t'))\cr & = 2 k_B
T\eta \delta (t-t')\cr}\eqno{(4c)}$$
To arrive at equations (3b) and the last term of (4c) we have
   assumed[13,14,15] Ohmic spectral density for the bath oscillators, i.e.,
$\rho(\omega)={\pi\over 2} \sum_{\alpha}{\lambda^2_{\alpha}\over
m_{\alpha}\omega_{\alpha}} \delta(\omega-\omega_c) =\eta \omega
e^{-\omega/\omega_c}.$
The upper cut-off frequency $\omega_c$ is assumed to be much
larger than the characteristic frequencies of the system. The
equations (3b) and (4c) correspond to the well known Markovian
limit and are valid for time scales $t>1/\omega_c$, which can be
made arbitrarily small by appropriately choosing $\omega_c$. For
details we refer to [14,15].
It should be noted that the transient terms have been neglected at time scales
$t>\omega_c^{-1}$ to arrive at (3b), and is
perfectly valid under Markovian approximation[15]. It is to be
noticed that $A'(Q)=$constant corresponds to a uniform friction
coefficient. Redefining,
$\eta [A'(Q)]^2=\eta (Q)$ and ${f(t)\over{\sqrt{ T\eta}}}
   \longrightarrow f(t),$
and putting $M=1$, we get,
$$\dot Q = P,\eqno{(5a)}$$
$$\dot P = - V'(Q) - \eta(Q) P+\sqrt{\eta(Q)T} f(t),\eqno{(5b)}$$
with
$$ < f(t) f(t') > = 2 k_B\delta (t-t').\eqno{(5c)}$$
It is instructive to note that one could take $\eta(Q)$ to be
constant piecewise along $Q$; in each piece of these $Q$ segments (5b) would
correspond to a constant friction coefficient but with the same
statistical character of $f(t)$ as in any other $Q$ intervals.

\vskip .3 true cm
\noindent{\bf 3. Microscopic Markovian Langevin equation with
space dependent friction and temperature}
\vskip .2 true cm
	We have so far derived the Langevin equation of motion
(from a microscopic Hamiltonian) of a Brownian particle with
space dependent friction keeping the temperature constant. We,
now, consider a system for which the temperature too is space
dependent $T(Q)$. At this point it is pertinent to note the
following
important fact, however. It is quite well known that when a
charged Brownian particle is subjected to an electric field charge
current results. Similarly, when it is subjected to a thermal
gradient thermal current flows in the system. However, in the former case
the effect of the electric field can be incorporated in the
particle Hamiltonian as a potential term whereas temperature
gradient cannot be incorporated as the potential term in the
Hamiltonian formalism.  Therefore, in order to incorporate the
effect of the temperature inhomogeneity we reason as follows. The
Brownian particle comes in contact with a continuous sequence
of independent  temperature
baths as its coordinate $Q$ changes in time. Equivalently, each
space point of the system is in equilibrium with a thermal bath
of characteristic temperature $T(Q)$. In what follows we
accept this idea and incorporate temperature inhomogeneity
into the equations of
motion (5). Henceforth, for notational simplicity, the
coordinate $Q$ and momentum $P$ are replaced by the corresponding
lower case letters $q$ and $p$, respectively, reserving $P$ for
probability distribution.

	For the sake of argument we consider, for the time
being, the system to be subdivided in space $q$ into several small segments
and represent the segments $\Delta q$ around $q$ by indices $i$.
Each segment is
connected to an independent thermal bath at temperature $T_i$
with corresponding random forces $f_i(t)$. The last term
on the right hand side of eq. (5b), is therefore
replaced by $\sqrt{\eta (q)T_i}f_i(t)$ for the segment $i$. As
the two different segments are each coupled to an independent
temperature bath we have $<f_i(t) f_j(t') > = 2 k_B\delta_{ij}\delta (t-t')$.
Because $f(t)$ is $\delta$-correlated in time, as the particle
evolves dynamically the fluctuation force
$f_i(t)$ experienced by the Brownian particle while in the space
segment $i$ at time $t$ will have no memory about the fluctuating force
experienced by it at some previous time $t'$ while in the space
segment $j\not =i$. Hence the space-dependent index $i$ in
$f_i(t)$ can be ignored. Now, taking a continuum limit, the
stochastic equations of motion of the Brownian particle, in an
inhomogeneous medium with space dependent friction and
nonuniform temperature, are given by,
$$\dot q = p,\eqno{(6a)}$$
and
$$\dot p = - V'(q) -\eta (q)p+\sqrt{\eta(q)T(q)} f(t),\eqno{(6b)}$$
with
$$ < f(t) f(t') > = 2 k_B \delta(t-t').\eqno{(6c)}$$
It is also important to note and worth repeating that as long as
the random force is
delta correlated in time, the final results remain unaffected
provided we incorporate space dependence in $f(t)\longrightarrow
f(q,t)$ such that $<f(q,t)f(q',t)>=2g(q-q')\delta(t-t')$ with $g(0)=1$.
\vskip .3 true cm
\noindent{\bf 4. Derivation of Kramers and Smoluchowski equations}
\vskip .2 true cm
	It is, now, a straightforward exercise to derive the
corresponding Fokker-Planck equation. We put
$M=1$ so that $p=v$, the velocity of the Brownian particle. The
stochastic differential equations (6a) and (6b) can be converted
into an equation for probability density $P(q,v,t)$ using the
well-known van Kampen lemma[16]. To this end, we consider a cloud
of initial "phase points" of density $\rho (q,v,t)$ in $(q,v)$
"phase space" each point $(q,v)$ of which are evolving in time
according to equations (6a) and (6b). The phase fluid
evolves according to the stochastic
Liouville equation (continuity equation)

$${\partial \rho\over\partial t} = -\nabla_q\cdot (\dot
q\rho)-\nabla_v \cdot (\dot v\rho).\eqno{(7)}$$
In order to obtain the equation for the evolution of $P(q,v,t)$ we
ensemble average ($<...>$) eq. (7) over all realizations of the random
force of given statistics and use the well-known result (van
Kampen lemma)[16]
$$< \rho > = P(q,v,t).\eqno{(8)}$$
The averaging procedure is carried out after substituting for
$\dot q$ and
$\dot v$ in eq. (7) from eq. (6). A term like $<\rho f(t)>$
appears which is evaluated by using the Novikov theorem[17]. For
details see reference [18,19]. From this we obtain the desired
Fokker-Planck equation
$${\partial P(q,v,t)\over \partial t} = - v{\partial P(q,v,t)\over
\partial q} - V'(q) {\partial P(q,v,t)\over \partial v}$$
$$ +\eta (q){\partial\over\partial v} \{ v P(q,v,t) +k_B T(q)
{\partial\over\partial v} P(q,v,t)\}.\eqno{(9)}$$
This is the Kramers equation for space dependent friction
coefficient $\eta (q)$ and nonuniform temperature $T(q)$, derived from
microscopic theory. It should be noted that van Kampen had
assumed eq. (9) as the model Kramers equation to start
with to study the diffusion of a Brownian particle in a ring due
to the combined effect of space dependent friction coefficient
and the temperature inhomogeneity[6]. Equations (6) and (9) are valid
for all friction coefficients, low as well as high. It is,
however, hard to solve eq. (9) in general cases. Moreover, in
many of the practical situations one does not need the detailed
motion of the Brownian particle at time scales much smaller than
the characteristic time scales of order $\eta^{-1}$. Therefore,
sometimes it is unnecessary to retain the fast variables $v$.

	In most of the problems of physical interest (overdamped
case) the marginal distribution $P(q,t)$ suffices to describe the
motion of a Brownian particle. In the case of uniform systems,
that is, when $\eta=$constant and $T=$constant, the reduction of
$P(q,v,t)$ to $P(q,t)$ is well known and goes by the name of
adiabatic elimination. One simply sets $\dot p=0$ in (6b) to obtain the
overdamped Langevin equation. From there one obtains the
Fokker-Planck equation for $P(q,t)$. The overdamped Langevin
equation so obtained is correct to order $\eta^{-1}$. For
inhomogeneous systems, however, the integration of (7) to obtain
the equation for $P(q,t)$ is not easy. Moreover, simply ignoring the
$\dot p$
term in (6b) is not correct and leads to unphysical results. For
instance, the resulting marginal distribution function so
obtained does not conform to the correct equilibrium
distribution. However, Sancho, San Miguel and Duerr[20] have given a
systematic procedure to go over to the overdamped Langevin
equation for a system with space dependent friction coefficient
but at uniform temperature. The overdamped Langevin equation
obtained by Sancho et al. is correct to order $[\eta(q)]^{-1}$
and leads to physically valid equilibrium distribution function.

	Following the prescription of Sancho et al.[20], we obtain
the overdamped Langevin equation for an inhomogeneous system
with nonuniform temperature field $T(q)$ and is given as
$$\dot q = -{V'(q)\over \eta(q)} - {1\over 2[\eta (q)]^2}\{
T(q)\eta'(q)+\eta(q)T'(q)\}+\sqrt{{T(q)\over \eta(q)}} f(t),\eqno{(10)}$$
with
$$< f(t)f(t')> = 2 k_B \delta(t-t').$$
The corresponding Fokker-Planck equation for the overdamped case
(the Smoluchowski equation), with $k_B$ set equal to 1, is
$${\partial P(q,t)\over \partial t} ={\partial\over\partial
q}{1\over \eta(q)}\bigg [ {\partial\over\partial q}T(q) P(q,t)+
V'(q) P(q,t)\bigg ].\eqno{(11)}$$

	As pointed out earlier by van Kampen, the diffusion
equation (11) in the absence of external potential has neither the form
${\partial P\over \partial t} = {\partial^2\over\partial q^2} D(q) P(q),$
nor  ${\partial P\over \partial t} = {\partial\over\partial q}
D(q){\partial \over\partial q} P(q).$
It is clear that $T(q)$ and $\eta(q)$ influence the motion of the
Brownian particle in different ways and their combined effect
cannot be plugged together as the effect of an effective
diffusion coefficient $D(q)$. We note that our equation (11)
agrees with one of the forms obtained by van Kampen[6-8].

	As already mentioned earlier the friction coefficient
$\eta(q)$ only affects the relaxation process and not the
equilibrium distribution function in a constant temperature
field. Given enough time the system finds its equilibrium state.
In contrast, the case of nonuniform temperature field changes
the concept of steady states. It can be
readily verified that if the external potential is unbounded at
infinity, i.e., $V(q)\longrightarrow \infty$ as
$q\longrightarrow \pm \infty$, then the system evolves to
a steady state $P_s(q)$ obtained by setting the probability
current equal to zero.

$$P_s(q,t) = {C\over T(q)} e^{-\int^q {V'(q)\over T(q')}dq'}\eqno{(12)}$$
where $C$ is a normalization constant. The solution in no way
resembles the distribution decided by the usual Boltzmann factor
alone. $P_s$ is not a local function of $V(q)$. The nonlocal
dependence of $P_s(q)$ on $T(q)$ and $V(q)$ forces the "relative
stability" of the system in two different local minima to depend
sensitively on the temperature profile along the entire pathway
connecting the two minima[2-5]. Moreover for particular choices of
$T(q)$ it may so happen that $P_s(q)$ may show extrema at
positions completely unrelated to the minima of the external
potential $V(q)$. Such a system with nonuniform temperature field
is inherently nonequilibrium and $P_s(q)$ describes distribution
of nonequilibrium steady states.

\vskip .3 true cm
\noindent{\bf 5. Conclusion}

	We have given a systematic microscopic derivation of
Kramers equation of motion of a Brownian particle in a medium
where friction coefficient is space dependent and having
nonuniform temperature. We
further obtain the Smoluchowski limit of the Kramers equation
following the procedure given by Sancho, et al.[20]. We thus arrive
at the correct overdamped Langevin equation for such a system.
The microscopic treatment followed in this work helps resolve
the controversy regarding the correct form of
the diffusion equation followed by a Brownian particle in an
inhomogeneous medium. We argue that the microscopic derivation
of the equations makes their application to systems such as the
thermal ratchets self-consistent[10-12]. Moreover, in many cases
the numerical solution of the Langevin equation is much more
transparent to appreciate physically, and the derived overdamped
Langevin equation could thus be of some practical use.
\vskip .3 true cm
\noindent{\bf Aknowledgements}
\vskip .2 true cm
	One of us(AMJ) thanks Professor N. Kumar for continued
discussions on this subject and related problems.
\vfill
\eject
\vskip .3 true cm
\noindent{\bf References}
\vskip .2 true cm
\item{[1]} N.G. van Kampen, {\it Stochastic Processes in
Physics and Chemistry}(North Holland, Amsterdam, 1981); C.W.
Gardiner, {\it Handbook of Stochastic Methods}(Springer
Verlag, Berlin, 1983).
\item{[2]} R. Landauer, Helv. Phys. Acta {\bf 56}, 847(1983);
J. Stat. Phys. {\bf 53}, 233(1988), and references therein.
\item{[3]} R. Landauer, Physica A{\bf194}, 551(1993), and
references therein.
\item{[4]} R. Landauer, Physics Today {\bf 31}, 23(November 1978).
\item{[5]} R. Landauer, Ann. N.Y. Acad. Sc. {\bf 316},
433(1979).
\item{[6]} N.G. van Kampen, IBM J. Res. Develop. {\bf 32},
107(1988).
\item{[7]} N.G. van Kampen, Z. Phys. B{\bf 68}, 135(1987).
\item{[8]} N.G. van Kampen, J. Maths. Phys. {\bf 29},
1220(1988).
\item{[9]} M. Buettiker, Z. Phys. B{\bf 68}, 161(1987).
\item{[10]} M.M. Millonas, Phys. Rev. Lett. {\bf 74}, 10(1995),
and references therein.
\item{[11]} A.M. Jayannavar and M.C. Mahato, unpublished.
\item{[12]} A.M. Jayannavar, unpublished.
\item{[13]} A.O. Caldeira and A.J. Leggett, Physica A{\bf 121},
587(1983); Annals. Phys. {\bf 149}, 374(1983).
\item{[14]} K. Lindenberg and V. Seshadri, Physica A{\bf 109},
483(1981).
\item{[15]} A.M. Jayannavar, Z. Phys. B{\bf 82}, 153(1991).
\item{[16]} N.G. van Kampen, Phys. Rep. C{\bf 24}, 172(1976).
\item{[17]} E.A. Novikov, Zh. Eksp. Teor. Fiz. {\bf 47},
1919(1964)[Sov. Phys. JETP {\bf 20}, 1290(1965)].
\item{[18]} A.M. Jayannavar and N. Kumar, Phys. Rev. Lett. {\bf
48}, 553(1982).
\item{[19]} A.M. Jayannavar, Phys. Rev. E{\bf 48}, 837(1993).
\item{[20]} J.M. Sancho, M. San Miguel and D. Duerr, J. Stat.
Phys. {\bf 28}, 291(1982).
\vfill
\eject
\end